\documentclass[twocolumn]{aastex62}

\newcommand\teff{\mbox{$T_\mathrm{eff}$}}
\newcommand\lya{Ly$\alpha$}
\usepackage{color}

\accepted{\today}
\submitjournal{ApJ}

\begin{document}

\title{Lyman-$\alpha$ Observations of High Radial Velocity Low-Mass Stars Ross 1044 and Ross 825}

\correspondingauthor{Adam C. Schneider}
\email{aschneid10@gmail.com}

\author{Adam C. Schneider}
\affil{School of Earth and Space Exploration, Arizona State University, Tempe, AZ, 85282, USA}

\author{Evgenya L. Shkolnik}
\affil{School of Earth and Space Exploration, Arizona State University, Tempe, AZ, 85282, USA}

\author{Travis S. Barman}
\affil{Lunar \& Planetary Laboratory, University of Arizona, 1629 E. University Blvd, Tucson, AZ 85721, USA}

\author{R. Parke Loyd}
\affil{School of Earth and Space Exploration, Arizona State University, Tempe, AZ, 85282, USA}

\begin{abstract}

The discovery of habitable zone (HZ) planets around low-mass stars has highlighted the need for a comprehensive understanding of the radiation environments in which such planets reside. Of particular importance is knowledge of the far-ultraviolet (FUV) radiation, as low-mass stars are typically much more active than solar-type stars and the proximity of their HZs can be one tenth the distance. The vast majority of the flux emitted by low-mass stars at FUV wavelengths occurs in the Lyman-$\alpha$ line at 1216 \AA. However, measuring a low-mass star's Lyman-$\alpha$ emission directly is almost always impossible because of the contaminating effects of interstellar hydrogen and geocoronal airglow.  We observed Ross 825 (K3) and Ross 1044 (M0), two stars with exceptional radial velocities, with the STIS spectrograph aboard the {\it Hubble Space Telescope (HST)}.  Their radial velocities resulted in significant line shifts, allowing for a more complete view of their Lyman-$\alpha$ line profiles.  We provide an updated relation between effective temperature and Lyman-$\alpha$ flux using {\it Gaia} DR2 astrometry as well as updated, model-independent relationships between Lyman-$\alpha$ flux and UV flux measurements from the {\it Galaxy Evolution Explorer (GALEX)} for low-mass stars.  These new relations, in combination with {\it GALEX}'s considerable spatial coverage, provide substantial predictive power for the Lyman-$\alpha$ environments for thousands of nearby, low-mass stars. 

\end{abstract}


\section{Introduction}

Both K and M type stars have distinct advantages for studying exoplanets.  M dwarfs are the most abundant stars in the Galaxy \citep{boch10}, have large planet occurrence rates \citep{dress15, hard19, tuomi19}, have numerous observational advantages for exoplanet characterization, and have therefore emerged as targets for current and future exoplanet investigations.  K dwarfs, while not as numerous as M dwarfs, also have substantial planet occurrence rates \citep{mul15} and the reduced X-ray and UV radiation incident on HZ planet atmospheres around K dwarfs has led some to suggest that K dwarfs might be the most promising targets for the detection of biology outside of our solar system \citep{hell14, cuntz16, rich19, arney19}.  For these reasons, understanding the properties of both K and M dwarfs is an essential component of current and upcoming planet-hunting missions such as TESS and PLATO \citep{rauer14,rick15}. 

The UV radiation emitted by low-mass stars can chemically modify, ionize, and even erode a planetary atmosphere over time, drastically affecting their habitability \citep{kast93, licht10, seg10, hu12, lug15}. Furthermore, several studies have shown that incident UV flux could lead to the formation of abiotic oxygen and ozone, resulting in potential false-positive biosignatures \citep{dom14, tian14, har15}.  The most prominent FUV emission line, Lyman-$\alpha$ (\lya; $\lambda$1215.67 \AA), comprises $\sim$37\%--75\% of the total 1150--3100 \AA\ flux from most late-type stars \citep{france13}.  Because \lya\ can affect the photodissociation of important molecules such as H$_2$O and CH$_4$, any photochemical models assessing potential biosignatures or atmospheric abundances will require accurate \lya\ host star flux estimates (e.g., \citealt{rug15}).         

In almost all cases, \lya\ cannot be directly observed because of 1) attenuation of photons by optically thick Hydrogen absorption located in the intervening interstellar medium (ISM) and 2) contamination from Earth's own \lya\ geocoronal airglow.  In fact, there exists only one complete \lya\ spectrum in the {\it HST} archive, that of Kapteyn's star (an M1 subdwarf; \citealt{gui16}).  With radial velocity (RV) of 245 km s$^{-1}$, its stellar \lya\ emission feature is shifted out of the geocoronal line core and ISM absorption.  

For all other stars, \lya\ emission may only be estimated.  \cite{wood05} reconstructed \lya\ fluxes by determining interstellar Hydrogen column densities and velocities inferred from deuterium and metal lines.  Alternatively, reconstructions have been performed by fitting 1 or 2 Gaussians to what remains of the wings of the \lya\ profile \citep{france12, young16, young17}.  Such reconstructions have been used extensively to produce correlations between \lya\ fluxes and other spectral emission lines \citep{linsky13} and broadband UV photometry from {\it GALEX} \citep{shk14b}. These correlations have been used to evaluate the capabilities of different spectral types to develop and sustain life (e.g., \citealt{cuntz16}).  

The lack of observational \lya\ constraints on low-mass stars has limited progress in predicting the environments that surround them.  In an effort to further inform models of low-mass stellar UV flux levels, we have extended the observational sample of stars for which the shape and majority of \lya\ flux can be measured beyond just Kapteyn's star. We identified two low-mass targets, Ross 825 (K3) and Ross 1044 (M2).  

\begin{deluxetable*}{lcccccccc}
\tablecaption{Target Sample}
\tablehead{
\colhead{Property} & \colhead{Ross 825} & \colhead{Ref.} & \colhead{Ross 1044} & \colhead{Ref.} &  }
\startdata
2MASS Name & 21111696$+$3331272 & 1 & 15032457$+$0346574 & 1  \\
Spectral Type & K3 & 2 & M0 & 3  \\
$\pi$ (mas) & 10.17$\pm$0.08 & 4 & 26.9 $\pm$8.4 & 5  \\
$\mu_{\alpha}$ (mas yr$^{-1}$) & 502.63 $\pm$0.17 & 4 & -899.1$\pm$4.7 & 5  \\
$\mu_{\delta}$ (mas yr$^{-1}$) & 159.31 $\pm$0.29 & 4 & 695.3$\pm$4.6 & 5  \\
RV (km s$^{-1}$) & -340.17$\pm$0.67 & 4 & -169.55$\pm$1.80  & 4  \\
Age (Gyr) & $>$10 & 3 & $>$10 & 3 \\
\enddata
\tablerefs{(1) 2MASS \cite{skrut06}; (2) \cite{bid85}; (3) This work; (4) \cite{gaia16, gaia18}; (5) \cite{finch16}}
\end{deluxetable*}

\section{Target Sample}

Targets for this program were chosen based on their spectral types (later than K0), distances (within 100 pc), and large absolute radial velocities.  We chose a radial velocity threshold of 150 km s$^{-1}$, which results in a $\approx$0.6 \AA\ wavelength shift of the \lya\ emission line.  Such stars are rare because the vast majority of low-mass stars in the solar neighborhood have absolute radial velocities less than 50 km s$^{-1}$.  \cite{shk12} found an average radial velocity of 1.6 km s$^{-1}$ with a standard deviation of 23.9 km s$^{-1}$ for a sample of 159 M dwarfs within 25 pc.  

We identified two targets, Ross 825 and Ross 1044, that satisfied these criteria and were bright enough to acquire a suitable signal-to-noise ratio (S/N) in a reasonable amount of {\it HST} time.  The properties of these two stars are summarized below.

\subsection{Ross 825}

Ross 825 was first identified as a high proper motion star by \cite{ross29}, who estimated a total proper motion of 0\farcs48 yr$^{-1}$.  \cite{bid85} compiled spectral classifications determined by G.~P.~Kuiper and published a spectral type of K3 for Ross 825.  Ross 825 was also noted as a close visual double in that work.  The radial velocity of Ross 825 was first measured in \cite{carney87}, who determined a value of -340.68 km s$^{-1}$.

Gaia observations \citep{gaia16, gaia18} of Ross 825 confirmed the binary nature of this source (separation $\approx$1\farcs6) and its radial velocity (-340.16$\pm$0.67 km s$^{-1}$).  It is also resolved as a binary in the STIS acquisition images acquired through this investigation. \cite{stass18} provided an effective temperature estimate ($T_{\rm eff}$) of 4680$\pm$177 K, which corresponds to a spectral type between K3 and K4 \citep{pec13}, consistent with its measured spectral type.  \cite{sch06} provided Stromgren $ubvy$--$\beta$ photometry of Ross 825 and determined a photometric metallicity of [Fe/H] = -1.28 based on empirical relations.  An [Fe/H] value of -1.28 implies an age $>$10 Gyr according to the models of \cite{dotter17}, which is supported by its large absolute radial velocity.  Relevant properties of Ross 825 are summarized in Table 1. 

\subsection{Ross 1044}

\cite{ross39} identified Ross 1044 as a high proper motion star with a total proper motion of 1\farcs27 yr$^{-1}$.  A spectral type of K7 was first published for this star in \cite{bid85}.  \cite{newt14} determined a near-infrared spectral type of M2 for Ross 1044.  We determine a spectral type of M0 by comparing the publicly available optical spectrum from the Palomar/MSU survey \citep{reid95} to SDSS standards \citep{boch10}.  Ross 1044 was found to have a large absolute radial velocity in the Palomar/MSU nearby star spectroscopic survey (-159.1 km s$^{-1}$; \citealt{reid95}).  This radial velocity was confirmed in \cite{newt14} (-174$\pm$5 km s$^{-1}$) and {\it Gaia} DR2 (-169.55$\pm$1.80 km s$^{-1}$).   Ross 1044 was also found to have a low metallicity ([Fe/H] = -1.01$\pm$0.21; \citealt{newt14}), which is consistent with an age $>$10 Gyr \citep{dotter17}. This star also showed no clear signature of H$\alpha$ either in emission or absorption in the optical spectrum from the Palomar/MSU survey \citep{reid95}.  Note that while Ross 1044 has a measured radial velocity in {\it Gaia} DR2, there is no parallax available.  This is likely because Ross 1044 has a {\tt\string visibility\_periods\_used} value of 5 in the {\it Gaia} DR2 catalog, which does not satisfy the criterion needed for a five-parameter solution in equation 11 of \cite{lind18}. Two parallax measurements exist in the literature: 47.0$\pm$4.1 mas \citep{van95} and 26.9$\pm$8.4 mas \citep{finch16}.  The ``quality of interagreement'' is listed as ``Poor'' in \cite{van95}, and we therefore adopt the \cite{finch16} parallax for this work.  Relevant properties of Ross 1044 are summarized in Table 1.  

\begin{figure*}
\plotone{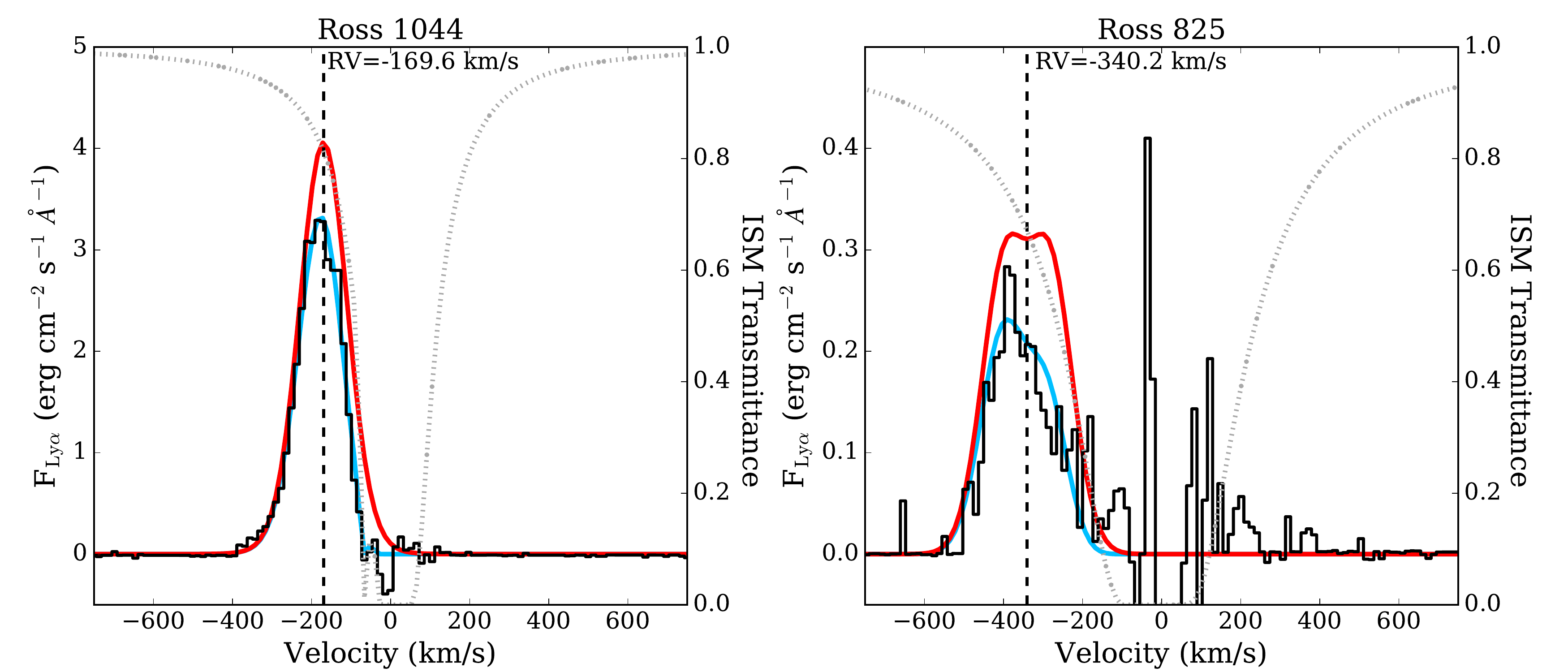}
\caption{HST/STIS spectra (black lines) for Ross 1044 (left) and Ross 825 (right).  The best fitting model is show in blue, with the resulting intrinsic \lya\ flux shown in red.  The ISM transmittance curve is shown as a dotted line.}  
\end{figure*}

\section{Observations}

We observed Ross 825 and Ross 1044 during {\it HST} Cycle 25 with the Space Telescope Imaging Spectrograph (STIS) aboard {\it HST} (PID \#15286).   For these observations, STIS is preferred over the Cosmic Origins Spectrograph (COS) because COS is a slitless spectrograph, which results in heavy contamination in the \lya\ region from geocoronal emission.  We employed the G140M grating (R $\sim$11400) with a 52\arcsec $\times$ 0\farcs1 slit.  For each observation, we set the central wavelength to 1222 \AA.  Observing dates and total exposure times are provided in Table 2.  

For several of the visits for each object, the {\it HST} $calstis$ pipeline failed to locate the correct location of the star.  For both stars, we determined trace positions manually and reran the $calstis$ pipeline for every visit.  For Ross 825, one visit was especially noisy and a trace position could not be determined.  These data are not included in the final spectrum.  Final spectra were produced by coadding the spectra for each visit together for each star and are shown in Figure 1.  While the binary nature of Ross 825 was seen in the STIS acquisition images, the \lya\ spectrum comes solely from the primary component of this system.  All subsequent analysis of Ross 825 corresponds exclusively to the primary object of this pair.

\begin{deluxetable}{lcccccccc}
\tablecaption{HST/STIS Observations}
\tablehead{
\colhead{Object} & \colhead{Obs.\ Dates} & \colhead{Exposure Time} \\
 & \colhead{UTC} & \colhead{(s)}  }
\startdata
Ross 825 & 30 July 2018, 3 Aug. 2018 & 17079  \\
Ross 1044 & 26 Nov. 2018, 30 Nov. 2018 & 15186 \\
\enddata
\end{deluxetable}

\section{Analysis}

\subsection{\lya\ Line Profiles}

For both Ross 825 and Ross 1044, the majority of their \lya\ emission is directly measured with our STIS observations.  However, neither target is immune to intervening ISM absorption.  For this reason, ISM modeling is necessary to retrieve the intrinsic \lya\ flux for both stars.  Following \cite{bour15}, we model the ISM using Voigt profiles for both atomic hydrogen and deuterium.  We fix the D {\footnotesize I}/ H {\footnotesize I} ratio to 1.5 $\times$ 10$^{-5}$ (\citealt{heb03,linsky06}).  We fix the radial velocity of the \lya\ emission to the stellar radial velocity measurements given in Table 1.  Free parameters then include the intrinsic \lya\ flux, the column density of hydrogen, and the radial velocity of the intervening ISM.  While there may be multiple ISM components along the line of sight to each star, we model them as a single component.

Whether or not intrinsic \lya\ line profiles contain a self-reversal, making them single- or double-peaked, is a large source of uncertainty for \lya\ reconstructions, leading to flux differences as large as 30\% \citep{young16}.  For strong lines, such as \lya, the emission in the wings and the core of the line can form in different atmospheric layers, leading to self-reversals in the core \citep{peacock19}.  While the Sun shows a self-reversal in its \lya\ core, \cite{young16} argue that M dwarf \lya\ profiles should be single-peaked because M dwarf Mg {\footnotesize II} profiles show signs of self-absorption.  However, we note that the presence of reversals in observed Mg {\footnotesize II} line cores of M dwarfs is also a consequence of spectral resolution.  \cite{font16} presented a STIS/E230H spectrum of the M2 star GJ 832, which shows clear signs of reversals in its Mg {\footnotesize II} line cores.  These reversals are not seen in the MUSCLES NUV spectra (see Figure 5 of \citealt{france16}), which were taken with the COS/G230L setting, which has $\sim$1 order of magnitude less spectral resolution than STIS/E230H.

Our STIS spectra have resolutions similar to COS/G230L.  All models are convolved with the STIS line spread function to compare to observations.  For Ross 1044, the \lya\ line shows a clear Gaussian shape, and the peak of the \lya\ emission occurs at the same RV as the stellar RV measurement.  Ross 825, on the other hand, has \lya\ emission peaking approximately 60 km s$^{-1}$ away from the measured stellar RV, suggestive of a double-peaked profile.  We therefore model our M dwarf target (Ross 1044) without a self-reversal and explore models with and without self-reversals for our K dwarf target (Ross 825).  Note that a similar RV offset between peak \lya\ emission and the stellar RV is seen for Kepler-444 (spectral type = K1), which has a radial velocity of -121.4 km s$^{-1}$ \citep{dup16}. \cite{bour17b} found that the \lya\ emission for Kepler-444 is best fit by a double-peaked profile.    

For Ross 1044 we use a model consisting of two Gaussians, one strong and narrow and the other weaker and broad, following \cite{young16}.  We also fit the Ross 825 data with two Gaussians, though the second Gaussian in this case is subtracted from the first to account for the presence of a self-reversal in the line core.  We find posterior distributions for each model parameter using the {\tt\string emcee} package \citep{fm13}.  Fore each fit, we run 1000 walkers with 1000 steps, treating the first 300 steps as the burn in sample.  For both objects, we also performed a fit with a single Gaussian, and the two-Gaussian solution is preferred in both cases according to the Bayesian Information Criterion (BIC).  For Ross 825, we also attempted a three-Gaussian fit -- broad and narrow profiles as with Ross 1044 and a third component subtracted out.  Again, the BIC prefers the two-Gaussian model compared to the three-Gaussian case.  Our model fits imply that our STIS observed spectra recovered $\sim$74\% and $\sim$63\% of the total intrinsic \lya\ flux for Ross 1044 and Ross 825, respectively.  These percentages are significantly higher than that typically seen for field M dwarfs, which is usually $\lesssim$30\% \citep{young17}. The best fitting profiles are shown in Figure 1 with best fit parameters given in Table 3 and the resulting \lya\ fluxes are provided in Table 6. 

\begin{deluxetable*}{llrlrcccc}
\tablecaption{{\tt\string emcee} Best Fit Model Parameters}
\tablehead{
\colhead{} & \multicolumn{2}{c}{Ross 1044} & \multicolumn{2}{c}{Ross 825}\\
\cline{2-3}
\cline{4-5}
\colhead{Model Parameter} & \colhead{Prior} & \colhead{Value} & \colhead{Prior} & \colhead{Value}}
\startdata
Amplitude 1 (10$^{-14}$ ergs s$^{-1}$ cm$^{-2}$ \AA$^{-1}$) & $\mathcal{U}$(0,10) & 3.75$^{+0.34}_{-1.46}$ & $\mathcal{U}$(0,10) & 1.86$^{+1.82}_{-0.96}$  \\
Sigma 1 (km s$^{-1}$) & $\mathcal{U}$(0,5) & 0.24$\pm$0.01 & $\mathcal{U}$(0,10) & 0.27$^{+0.12}_{-0.02}$\\
Amplitude 2\tablenotemark{a} (10$^{-14}$ ergs s$^{-1}$ cm$^{-2}$ \AA$^{-1}$) & $\mathcal{U}$(0,5) & 1.55$^{+1.82}_{-0.96}$ & $\mathcal{U}$(0,5) & 0.31$^{+0.61}_{-0.20}$ \\
Sigma 2\tablenotemark{a} (km s$^{-1}$) & $\mathcal{U}$(0,25) & 0.35$^{+1.57}_{-0.12}$ & $\mathcal{U}$(0,5) & 0.24$^{+0.17}_{-0.03}$ \\
RV$_{\rm ISM}$ (km s$^{-1}$) &$\mathcal{U}$(-50,50) & 16.32$^{+5.18}_{-3.93}$ & $\mathcal{U}$(-50,50) & -13.39$^{+30.63}_{-24.77}$  \\
10$^{18}$ cm$^{-2}$ / d \tablenotemark{b} (pc) & $\mathcal{U}$(4,50) & 13.58$^{+3.28}_{-2.46}$ & $\mathcal{U}$(4,50) & 5.95$^{+2.72}_{-1.44}$ \\
\enddata
\tablenotetext{a}{The Gaussian described by Amplitude 2 and Sigma 2 was added to the first Gaussian in the case of Ross 1044, and subtracted in the case of Ross 825.}
\tablenotetext{b}{Range of H {\footnotesize I} column densities taken from \cite{redfield08}.}
\end{deluxetable*}

\section{Discussion}

\subsection{\lya\ Variability of Ross 1044}

Ross 1044 has sufficient S/N in each of the six visits to investigate for variability in the strength of the \lya\ line.  In Figure 2, we plot the the individual spectra from each visit, as well as our combined final spectrum.  While the general shape and flux is similar for each of the six visits, the total flux for the first three visits is less than that from the last three visits.  We quantify this by summing up the total flux in each visit, then normalizing to the total flux from the final combined spectrum, with uncertainties calculated in a Monte Carlo fashion.  We find that the total flux varied by $\sim$20\% between the smallest and largest flux values measured.  Note that there exists a STIS ``breathing'' effect, which can cause variability within {\it HST} orbits \citep{brown01, sing08}.  This effect should be mitigated by the use of spectra averaged over each visit.  The \lya\ variability of Ross 1044 is consistent with that seen in the \lya\ line of GJ 1132, where a variability up to 22\% is seen \citep{waalkes19} and Kepler-444, where variability of 20--40\% is seen between visits \citep{bour17b}.  This is also consistent with the variability seen in other UV spectral lines in \cite{loyd14}, where variability from 1--41\% is measured. The number and cadence of our observations prevents us from determining the cause of Ross 1044's variability (e.g., rotational modulations, flares).  A dedicated monitoring campaign could pinpoint the cause of this variability. 

 \begin{figure*}
\plotone{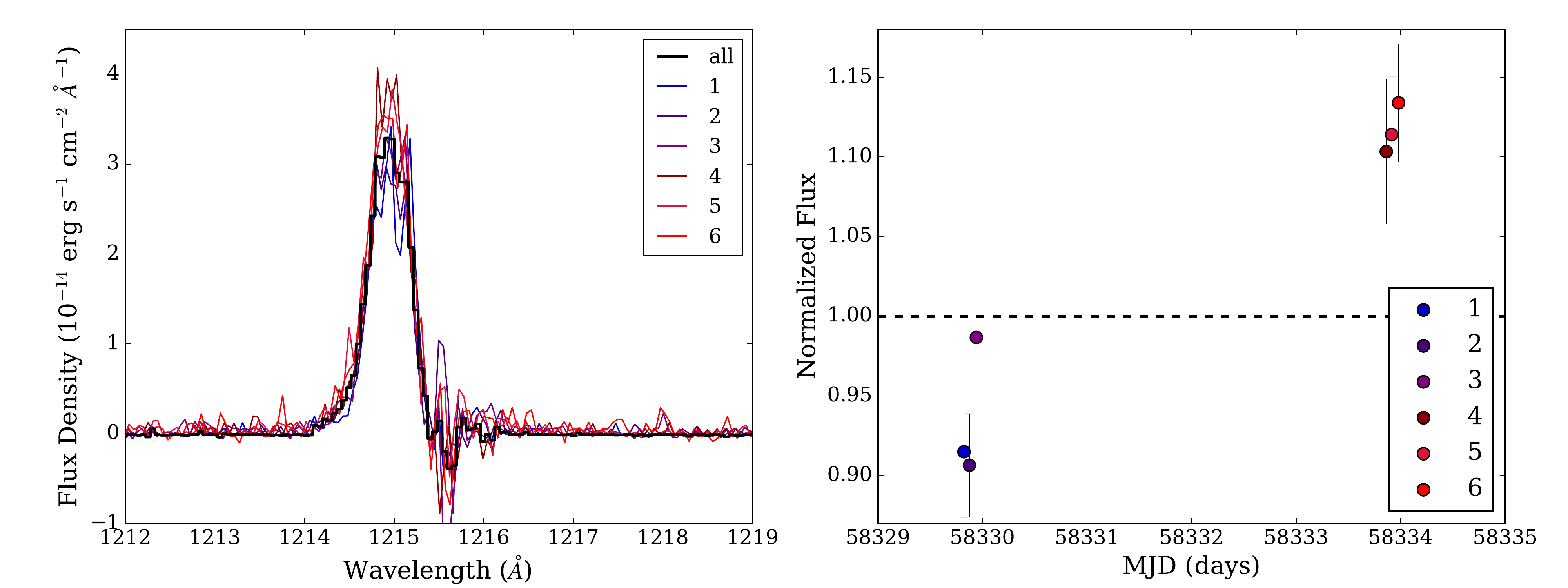}
\caption{{\it Left:} STIS \lya\ spectra of Ross 1044 for all 6 visits compared to the final coadded spectrum. {\it Right:} \lya\ flux variability of Ross 1044.  All points are normalized to the flux of the final coadded spectrum.}  
\end{figure*}

\subsection{\lya\ Correlations}
Correlating intrinsic \lya\ fluxes with stellar properties gives us insight into the behavior of high energy emission of stars in various evolutionary and temperature regimes.  Relationships between stellar properties and \lya\ flux values can also provide a powerful means to predict \lya\ fluxes for stars without \lya\ observations.  

\subsubsection{\lya\ Behavior with Effective Temperature}

\cite{linsky13} showed a correlation between stellar effective temperature (\teff) and \lya\ flux.  We revisit this relationship here to include additional measurements published since that paper as well as Ross 825 and Ross 1044.  These additions increase the entire sample size by $\sim$30\%, with 47 \lya\ measurements used in \cite{linsky13} and 61 used in this work, two of which are new measurements.  We also use updated \teff\ measurements as well as more precise parallaxes from Gaia DR2 \citep{gaia16,gaia18}. Table 6 lists these properties (\teff, age, parallax) and intrinsic \lya\ fluxes for all stars with \lya\ flux measurements.  For Kapteyn's Star, \cite{gui16} directly measured a \lya\ flux of 5.32 $\times$ 10$^{-13}$ erg s$^{-1}$ cm$^{-2}$, though no uncertainty is provided for this measurement.  We assume an uncertainty of 5\% for this directly measured line. 

Figure 3 shows the \lya\ flux for each star in Table 6 scaled to 1 AU versus their effective temperatures.  While \cite{linsky13} partitions different stellar samples based on their rotation rates, we instead choose to use age because age measurements are more readily available for this sample.  We note that determining precise stellar ages is a challenging process, and many different methods can be used depending on the star in question and the data available for that star.  All ages used in this work are provided in Table 6.  For more information on how a particular age was determined, please refer to the reference provided in the table. There is a clear difference between young ($<$1 Gyr) and old ($>$1 Gyr) samples in Figure 3.  \lya\ emission is typically much stronger from young stars than old stars with similar \teff\ values.  For old stars, a very clear trend of decreasing \lya\ flux strength with decreasing \teff\ can be seen.  We fit this trend using a least-squares optimization approach, with uncertainties handled in a Monte Carlo fashion.  We find:

\begin{equation}
{\rm log}_{10}(F_{\rm Ly\alpha})_{\rm 1 AU} = (6.754\pm0.263)\times10^{-4}(\teff) - {2.639\pm0.137} .
\end{equation}

The RMS scatter about the fit is $\sim$0.21 dex.  We also plot in Figure 3 the \lya\ flux scaled to the HZ for each target.  The distance to the HZ for each star is calculated using their effective temperatures and radii listed in Table 6 and equations 4 and 5 of \cite{kopp13a} and the updated coefficients in \cite{kopp13b}.  The HZ distance is taken as the average value between the distances to the moist greenhouse and maximum greenhouse boundaries.  While the relation between \lya\ flux and \teff\ flattens out for this sample, it is notable that the slope of this relation positive, implying that, in general, the \lya\ flux in the HZ increases with decreasing \teff.  The relationship between \lya\ HZ flux and \teff\ is:

\begin{equation}
{\rm log}_{10}(F_{\rm Ly\alpha})_{HZ} = (-1.601\pm0.243)\times10^{-4}(\teff) + {1.804\pm0.126} .
\end{equation}

The RMS scatter about the fit is $\sim$0.19 dex.

\begin{figure*}
\plotone{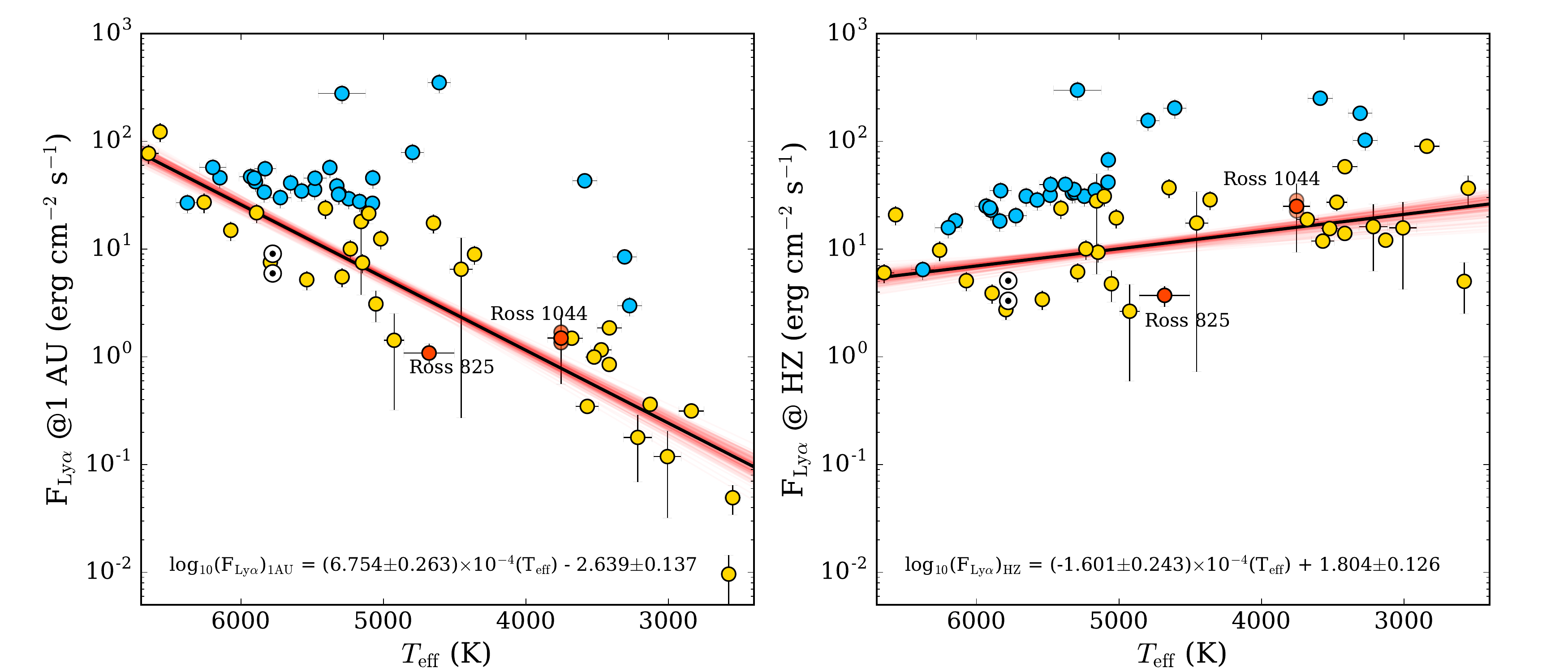}
\caption{{\it Left:} \lya\ Flux at 1 AU versus \teff\ for all objects in Table 6.  Young stars (age $<$ 1 Gyr) are plotted in light blue, while old stars (age $>$ 1 Gyr) are plotted in yellow. Ross 825 and Ross 1044 (age $>$ 10 Gyr) are plotted in red.  Two additional symbols are plotted for Ross 1044 representing the maximum and minimum variability determined in Section 5.1.  The two solar symbols represent values for the quiet and active Sun from \cite{linsky13}.  The solid black line represents the best fit to the data, with red lines showing individual fits from a Monte Carlo analysis. {\it Right:}  Same as the left panel with \lya\ fluxes scaled to the distance to the habitable zone for each star.}  
\end{figure*}

\subsubsection{Predicting \lya\ Fluxes with {\it GALEX} Photometry}

Because \lya\ emission is impossible to measure directly for almost all stars, and even indirect methods, such as reconstructions, require valuable {\it HST} time, correlations between \lya\ fluxes and other activity indicators are immensely valuable as predictive tools.  \cite{shk14b} used {\it GALEX} photometry and reconstructed \lya\ emission profiles from \cite{wood05}, \cite{france12,france13}, and \cite{linsky13} to show trends between NUV and FUV photometry and the reconstructed \lya\ fluxes for K and M dwarfs.  We update these correlations with precision parallaxes from {\it Gaia} and new and updated \lya\ fluxes from this work, \cite{gui16}, \cite{young16}, and \cite{young17}.  The new NUV relation contains three objects not included in the \cite{shk14b} relation; Ross 1044, Kapteyn's Star, and GJ 176.  We also produce these new correlations using flux at 1 AU instead of excess surface flux as in \cite{shk14b}, which avoids any uncertainties from model photospheric flux and radius estimates.  Thus for the NUV relation, we only include stars where the photospheric NUV flux is negligible, namely M stars (at all ages) and young K stars (\citealt{schneid18}, \citealt{rich19}).  

Table 4 summarizes all observed \lya\ flux measurements for M stars and young K stars with NUV detections in {\it GALEX}.  Note that there are many young K dwarfs from Table 4 that were observed by {\it GALEX}, however the majority of the sources were saturated (NUV $<$ 15 mag), and are thus not included here.  For objects that have detections from multiple {\it GALEX} surveys, the magnitudes presented in Tables 4 and 5 are the weighted average of their flux densities, converted back to magnitude.  Figure 4 shows a comparison of the the \lya\ flux values versus {\it GALEX} NUV flux values at 1 AU.  We perform a least-squares fit to the data and find the following relation between \lya\ flux and {\it GALEX} NUV fluxes: 

\begin{equation}
{\rm log}_{10}(F_{\rm Ly\alpha}) = (0.701\pm0.019)*{\rm log}_{10}(F_{\rm NUV}) + {0.193\pm0.043} .
\end{equation}

Uncertainties for the fit are determined in a Monte Carlo fashion.  The RMS scatter about this fit is 0.10 dex.

Some nearby, low-mass stars are too bright for reliable {\it GALEX} NUV measurements.  For these stars, it may be possible to use FUV fluxes.  We plot {\it GALEX} FUV fluxes versus \lya\ fluxes in Figure 4, with FUV data given in Table 5.  All K and M dwarfs with unsaturated {\it GALEX} FUV measurements are included in the FUV relation, with the total number of stars used increasing from 10 in \cite{shk14b} to 13 in this work with the addition of HD 97658, HD 40307, and GJ 176.  We perform a fit using the same methods as for {\it GALEX} NUV data and find:

\begin{equation}
{\rm log}_{10}(F_{\rm Ly\alpha}) = (0.742\pm0.038)*{\rm log}_{10}(F_{\rm FUV}) + {0.945\pm0.053} .
\end{equation}

The RMS scatter about this fit is 0.13 dex.  The utility of these correlations can be far-reaching, as {\it GALEX} has archived flux measurements for thousands of late-K and M dwarfs in the solar neighborhood (e.g., \citealt{ans15}, \citealt{jones16}, \citealt{kast17}, \citealt{miles17}, \citealt{schneid18}, \citealt{rich19}).

\begin{deluxetable*}{lrrrrrrrr}
\tablecaption{{\it GALEX} NUV and \lya\ Fluxes for M Dwarfs and Young K Dwarfs}
\tablehead{
\colhead{Name} & \colhead{Spectral} & \colhead{Age} & \colhead{plx\tablenotemark{a}} & \colhead{\lya} & \colhead{NUV} \\
\colhead{} & \colhead{Type} & \colhead{(Gyr)} & \colhead{(mas)} & \colhead{(erg cm$^{-2}$ s$^{-1}$)} & \colhead{(mag)}  }
\startdata
PW And & K2 & 0.15$^{+0.05}_{-0.02}$ & 35.29 $\pm$ 0.05 & (2.31$\pm$0.46)$\times$10$^{-12}$ & 15.317 $\pm$ 0.012 \\ 
Speedy Mic & K3 & 0.03$\pm$0.01 & 14.98 $\pm$ 0.27 & (1.85$\pm$0.37)$\times$10$^{-12}$ & 15.469 $\pm$ 0.009 \\ 
Ross 1044 & M0 & $>$10 & 26.9 $\pm$ 8.4 & (2.54$^{+0.14}_{-0.11}$)$\times$10$^{-14}$ & 21.571 $\pm$ 0.095 \\
Kapteyn's Star & sdM1 & 11.5$^{+0.5}_{-1.5}$ & 254.23 $\pm$ 0.03 & (5.27$\pm$0.26)$\times$10$^{-13}$ & 19.103 $\pm$ 0.048 \\
AU Mic & M1 & 0.02$\pm$0.01 & 102.83 $\pm$ 0.05 & (1.07$\pm$0.04)$\times$10$^{-11}$ & 15.614 $\pm$ 0.010 \\ 
GJ 832 & M2 & 8.4 & 201.41 $\pm$ 0.04 & (9.5$\pm$0.6)$\times$10$^{-13}$ & 18.393 $\pm$ 0.036 \\ 
GJ 176 & M2 & 4.0$\pm$0.3 & 105.56 $\pm$ 0.07 & (3.9$\pm$0.2)$\times$10$^{-13}$ & 18.825 $\pm$ 0.052 \\ 
GJ 436 & M3 & 4.2$\pm$0.3 & 102.50 $\pm$ 0.09 & (2.1$\pm$0.3)$\times$10$^{-13}$ & 20.685 $\pm$ 0.198 \\ 
AD Leo & M3 & 0.025$-$0.3 & 201.37 $\pm$ 0.07 & (8.07$\pm$0.20)$\times$10$^{-12}$ & 15.814 $\pm$ 0.017 \\ 
GJ 876 & M4 & 9.51$\pm$0.58 & 213.87 $\pm$ 0.08 & (3.9$\pm$0.4)$\times$10$^{-13}$ & 20.149 $\pm$ 0.087 \\ 
Prox Cen & M5.5 & 5.3$\pm$0.3 & 768.50 $\pm$ 0.20 & (4.37$\pm$0.07)$\times$10$^{-12}$ & 18.601 $\pm$ 0.086 \\ 
\enddata
\tablenotetext{a}{Parallaxes from {\it Gaia} \citep{gaia16, gaia18}, with the exception of Ross 1044 \citep{finch16}.}
\end{deluxetable*}

 \begin{figure*}
\plotone{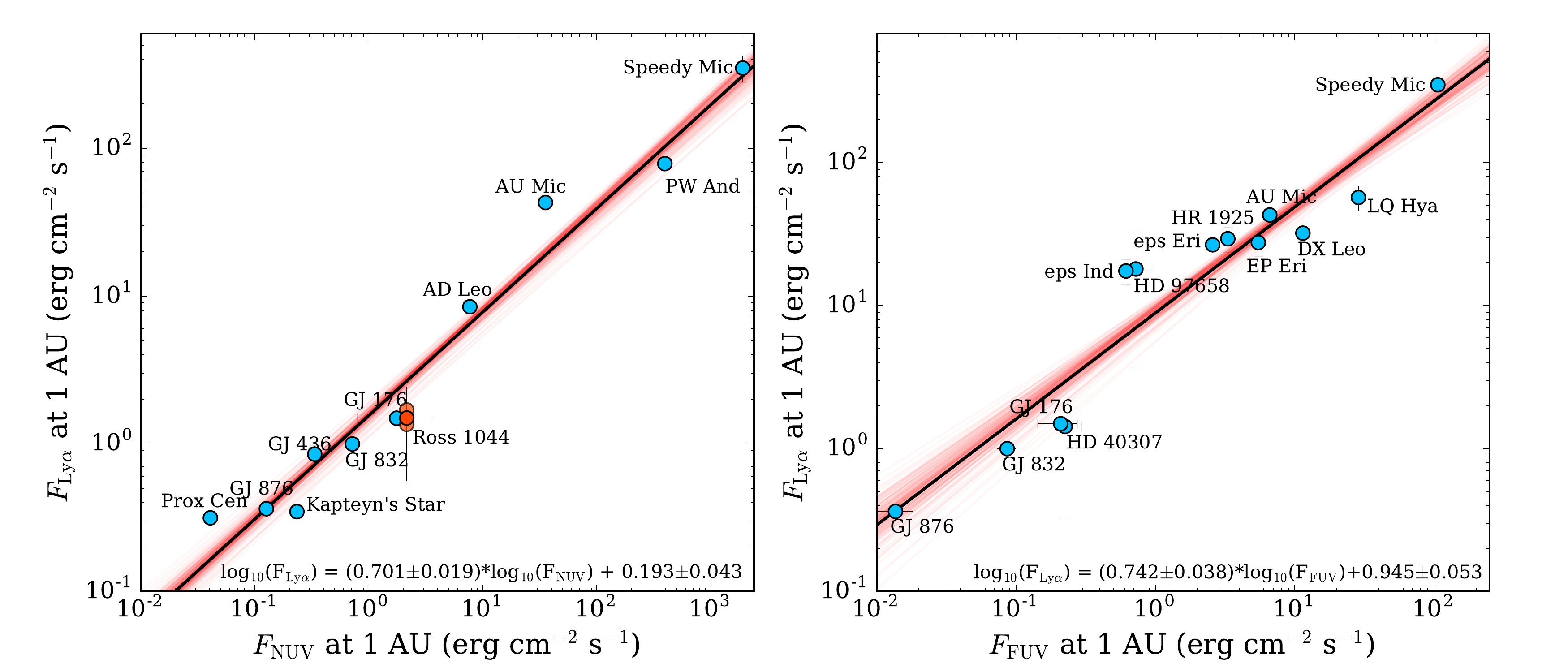}
\caption{{\it Left:} \lya\ flux versus {\it GALEX} NUV flux for low-mass stars (light blue symbols) and our new measurement for Ross 1044 (red symbol).  Two additional symbols are plotted for Ross 1044 representing the maximum and minimum variability determined in Section 5.1.  The solid black line represents the best fit to the data, with red lines showing individual fits from a Monte Carlo analysis. {\it Right:} Same as left panel for {\it GALEX} FUV.}  
\end{figure*}

\begin{deluxetable*}{lrrrrrrrr}
\tablecaption{{\it GALEX} FUV and \lya\ Fluxes for K and M Dwarfs}
\tablehead{
\colhead{Name} & \colhead{Spectral} & \colhead{Age} & \colhead{plx} & \colhead{\lya} & \colhead{FUV}\\
\colhead{} & \colhead{Type} & \colhead{(Gyr)} & \colhead{(mas)} & \colhead{(erg cm$^{-2}$ s$^{-1}$)} & \colhead{(mag)}  }
\startdata
DX Leo & K0 & 0.25$\pm$0.05 & 55.32 $\pm$ 0.06 & (2.31$\pm$0.46)$\times$10$^{-12}$ & 18.188 $\pm$ 0.053  \\ 
HR 1925 & K1 & 0.5$\pm$0.1 & 81.43 $\pm$ 0.05 & (4.57$\pm$0.91)$\times$10$^{-12}$ & 18.695 $\pm$ 0.091  \\ 
HD 97658 & K1 & 9.7$\pm$2.8 & 46.35 $\pm$ 0.05 & (9.1$^{+9.9}_{-4.5}$)$\times$10$^{-13}$ & 21.568 $\pm$ 0.308  \\ 
EP Eri & K2 & 0.2$\pm$0.1 & 96.54 $\pm$ 0.04 & (6.05$\pm$1.21)$\times$10$^{-12}$ & 17.779 $\pm$ 0.018  \\ 
$\epsilon$ Eri & K2 & 0.5$\pm$0.1 & 312.22 $\pm$ 0.47 & (6.1$\pm$0.2)$\times$10$^{-11}$ & 16.049 $\pm$ 0.023  \\ 
LQ Hya & K2 & 0.07$^{+0.03}_{-0.02}$ & 54.68 $\pm$ 0.07 & (4.01$\pm$0.80)$\times$10$^{-12}$ & 17.217 $\pm$ 0.037  \\ 
HD 40307 & K2.5 & 6.9$\pm$4.0 & 77.27 $\pm$ 0.03 & (2.0$^{+2.2}_{-0.9}$)$\times$10$^{-13}$ & 21.728 $\pm$ 0.346  \\ 
Speedy Mic & K3 & 0.03$\pm$0.01 & 14.98 $\pm$ 0.27 & (1.85$\pm$0.37)$\times$10$^{-12}$ & 18.604 $\pm$ 0.079  \\ 
$\epsilon$ Ind & K5 & 1.6$\pm$0.2 & 274.80 $\pm$ 0.25 & (3.10$\pm$0.62)$\times$10$^{-11}$ & 17.881 $\pm$ 0.047  \\ 
AU Mic & M1 & 0.02$\pm$0.01 & 102.83 $\pm$ 0.05 & (1.07$\pm$0.04)$\times$10$^{-11}$ & 17.436 $\pm$ 0.038  \\ 
GJ 832 & M2 & 8.4 & 201.41 $\pm$ 0.04 & (9.5$\pm$0.6)$\times$10$^{-13}$ & 20.687 $\pm$ 0.168  \\ 
GJ 176 & M2 & 4.0$\pm$0.3 & 105.56 $\pm$ 0.07 & (3.9$\pm$0.2)$\times$10$^{-13}$ & 21.130 $\pm$ 0.347  \\ 
GJ 876 & M4 & 9.51$\pm$0.58 & 213.87 $\pm$ 0.08 & (3.9$\pm$0.4)$\times$10$^{-13}$ & 22.560 $\pm$ 0.360  \\ 
\enddata
\end{deluxetable*}

\section{Conclusion}
We presented HST/STIS observations of Ross 1044 and Ross 825, two low-mass stars with sufficient to directly observe the majority of each star's intrinsic \lya\ flux ($\sim$74\% for Ross 1044 and $\sim$63\% for Ross 825).  We combined these new \lya\ measurements with updated astrometric information and physical parameters to explore the relationship between \lya\ flux and \teff.  We also constructed new correlations between \lya\ fluxes and {\it GALEX} UV photometry.

Astrometric missions such as {\it Gaia} will soon reveal numerous additional nearby stars with radial velocities for which such observations are possible.  High-RV targets are the only way to directly observe the majority of the \lya\ flux from low mass stars.  While the exact RV value needed depends on a target's distance and the ISM properties towards that target, a minimum absolute value of $\sim$100 km s$^{-1}$, corresponding to a wavelength shift of $\approx$0.4 \AA, is likely necessary.   Of the $\sim$73 thousand stars within 50 pc with radial velocity measurements in the {\it Gaia} DR2 catalog, only 120 (0.16\%) have radial velocities $>$100 km s$^{-1}$.  Each of these objects are thus invaluable as laboratories for investigating the UV environments of nearby stars.

\begin{longrotatetable}
\begin{deluxetable*}{rrrrrrrrrrrrrrr}
\tablecaption{Properties of All Stars with Intrinsic \lya\ Fluxes}
\tablehead{
\colhead{R.A.} & \colhead{Dec.} & \colhead{Other} & \colhead{Spectral} & \colhead{ref} & \colhead{$T_{\rm eff}$} & \colhead{ref} & \colhead{Radius} & \colhead{ref} & \colhead{plx\tablenotemark{a}} & \colhead{\lya} & \colhead{ref} & \colhead{Age} & \colhead{ref} \\
\colhead{($\degr$)} & \colhead{($\degr$)} & \colhead{Name} & \colhead{Type} & & \colhead{(K)} & & \colhead{$R_{\odot}$} & &  \colhead{(mas)} & \colhead{(erg cm$^{-2}$ s$^{-1}$)} & & \colhead{(Gyr)} &   }
\startdata
\cutinhead{New}
317.823653042 & $+$33.524935535 & Ross 825 & K3 & 1 & 4680$\pm$177 & 19 & 0.57$^{+0.15}_{-0.07}$ & 15 & 10.17$\pm$0.08 & (2.65$^{+0.38}_{-0.76}$)$\times$10$^{-15}$ & 3 & $>$10 & 3 \\
225.848498243 & $+$03.785579277 & Ross 1044 & M0 & 3 & 3754$\pm$95 & 25 & 0.380$\pm$0.029 & 25 & 26.9$\pm$8.4 & (2.54$^{+0.14}_{-0.11}$)$\times$10$^{-14}$ & 3 & $>$10 & 3 \\
\cutinhead{Old}
067.695503172 & $+$16.148537247 & SAO 93981 & F2 & 8 & 6567$\pm$49 & 26 & 1.474$^{+0.058}_{-0.047}$ & 27 & 22.12$\pm$0.06 & (1.41$\pm$0.28)$\times$10$^{-12}$ & 63 & 1.16$\pm$0.82 & 38 \\
\dots & \dots & Procyon & F5IV-V & 7 & 6648$\pm$71 & 21 & 2.13$\pm$0.07 & 19 & 284.56$\pm$1.26 & (1.47$\pm$0.03)$\times$10$^{-10}$ & 65 & 1.97$^{+0.19}_{-0.23}$ & 21 \\
066.577555282 & $+$21.470267340 & SAO 76609 & F8V & 9 & 6376$\pm$80 & 23 & 1.30$\pm$0.02 & 15 & 20.67$\pm$0.07 & (2.70$\pm$0.54)$\times$10$^{-13}$ & 63 & 0.63$\pm$0.05 & 41 \\
076.377476718 & $-$57.472196938 & $\zeta$ Dor & F9V & 4 & 6147$\pm$53 & 21 & 1.07$^{+0.02}_{-0.07}$ & 15 & 86.02$\pm$0.15 & (7.99$\pm$1.60)$\times$10$^{-12}$ & 63 & 0.68$\pm$0.47 & 38  \\
184.446711916 & $-$36.093975066 & HR 4657 & F9V & 4 & 6258$\pm$39 & 21 & 1.096$\pm$0.033 & 27 & 44.74$\pm$0.81 & (1.28$\pm$0.26)$\times$10$^{-12}$ & 63 & 1.8$\pm$0.5 & 37 \\
330.795021930 & $+$18.884241930 & V376 Peg & F9V & 9 & 6071$\pm$20 & 22 & 1.18$\pm$0.02 & 15 & 20.68$\pm$0.05 & (1.50$\pm$0.30)$\times$10$^{-13}$ & 63 & 3.83$^{+0.98}_{-0.70}$ & 22 \\
066.900016757 & $+$15.589088762 & V993 Tau & G0 & 8 & 6197$\pm$94 & 23 & 1.27$^{+0.03}_{-0.02}$ & 15 & 20.93$\pm$0.07 & (5.90$\pm$1.18)$\times$10$^{-13}$ & 63 & 0.63$\pm$0.05 & 41  \\
088.594850554 & $+$20.275864770 & $\chi$ Ori & G0V & 4 & 5898$\pm$25 & 17 & 0.9791$\pm$0.0091 & 17 & 113.12$\pm$0.32 & (1.27$\pm$0.25)$\times$10$^{-11}$ & 63 & 0.3$\pm$0.1 & 37 \\
238.171482022 & $+$42.454228071 & $\chi$ Her & G0 & 1 & 5890$\pm$53 & 28 & 1.7090$\pm$0.0200 & 29 & 63.16$\pm$0.15 & (2.04$\pm$0.41)$\times$10$^{-12}$ & 63 & 6.85$^{+0.42}_{-0.50}$ & 22 \\
271.599395673 & $-$36.019748261 & HR 6748 & G0V & 4 & 5932$\pm$80 & 23 & 0.98$^{+0.01}_{-0.04}$ & 15 & 58.13$\pm$0.12 & (3.73$\pm$0.75)$\times$10$^{-12}$ & 63 & 0.44$\pm$0.19 & 40 \\
\dots & \dots & Sun & G2V & \dots & 5778 & \dots & 1.0 & \dots & \dots & 5.95-9.15\tablenotemark{b} & 65 & 4.566$\pm$0.005 & 66 \\
\dots & \dots & $\alpha$ Cen A & G2V & 4 & 5793$\pm$7 & 17 & 1.2329$\pm$0.0037 & 17 & 754.81$\pm$4.11 & (1.01$\pm$0.20)$\times$10$^{-10}$ & 63 & 5.3$\pm$0.3 & 34 \\
112.676660493 & $-$37.339135172 & HR 2882 & G2V & 4 & 5830$\pm$76 & 22 & 0.93$^{+0.02}_{-0.01}$ & 15 & 45.93$\pm$0.03 & (2.76$\pm$0.55)$\times$10$^{-12}$ & 63 & 0.35$\pm$0.07 & 40  \\
168.133469825 & $+$35.813428317 & HR 4345 & G2 & 1 & 5906$\pm$29 & 28 & 0.99$\pm$0.03 & 30 & 44.14$\pm$0.04 & (2.09$\pm$0.42)$\times$10$^{-12}$ & 63 & 0.45$\pm$0.02 & 40 \\
049.841563537 & $+$03.370603059 & $\kappa$ Cet & G4 & 1 & 5723$\pm$76 & 17 & 0.9193$\pm$0.0247 & 17 & 109.34$\pm$0.31 & (8.44$\pm$1.70)$\times$10$^{-12}$ & 63 & 0.6$\pm$0.2 & 22 \\
129.458268090 & $-$06.806694022 & SAO 136111 & G5V & 7 & 5836$\pm$45 & 28 & 1.00$\pm$0.02 & 15 & 41.08$\pm$0.04 & (1.34$\pm$0.27)$\times$10$^{-12}$ & 63 & 0.51$\pm$0.14 & 40 \\
093.438506782 & $-$23.861458285 & HR 2225 & G6.5V & 4 & 5651$\pm$43 & 28 & 0.890$\pm$0.04 & 18 & 59.76$\pm$0.02 & (3.45$\pm$0.69)$\times$10$^{-12}$ & 63 & 0.28--0.36 & 39 \\
199.596454977 & $-$18.315774489 & 61 Vir & G7V & 4 & 5538$\pm$13 & 18 & 0.987$\pm$0.005 & 18 & 117.57$\pm$0.24 & (1.69$\pm$0.34)$\times$10$^{-12}$ & 63 & 9.41$^{+1.31}_{-3.15}$ & 22  \\
222.848032870 & $+$19.100275013 & $\xi$ Boo A & G7V & 7 & 5483$\pm$32 & 17 & 0.8627$\pm$0.0107 & 17 & 148.52$\pm$0.24 & (1.85$\pm$0.37)$\times$10$^{-11}$ & 63 & 0.2$\pm$0.1 & 37 \\
001.655138540 & $+$29.020738353 & HR 8 & G8V & 7 & 5327$\pm$39 & 17 & 0.9172$\pm$0.0090 & 17 & 72.58$\pm$0.05 & (4.77$\pm$0.95)$\times$10$^{-12}$ & 63 & 0.3$\pm$0.1 & 22 \\ 
220.129108915 & $-$16.209576638 & SAO 158720 & G8V & 4 & 5574$\pm$50 & 23 & 0.87$^{+0.02}_{-0.04}$ & 15 & 42.09$\pm$0.04 & (1.44$\pm$0.29)$\times$10$^{-12}$ & 63 & 0.62$\pm$0.07 & 40 \\
321.170945318 & $-$68.227107394 & SAO 254993 & G8V & 4 & 5480$\pm$80 & 23 & 0.87$\pm$0.01 & 15 & 48.06$\pm$0.34 & (2.47$\pm$0.49)$\times$10$^{-12}$ & 63 & 0.33$\pm$0.08 & 40 \\
026.009302877 & $-$15.933798651 & $\tau$ Cet & G8.5V & 4 & 5290$\pm$39 & 17 & 0.8154$\pm$0.0122 & 17 & 277.52$\pm$0.52 & (9.99$\pm$2.0)$\times$10$^{-12}$ & 63 & 5.6$\pm$1.2 & 37 \\
201.437998838 & $+$56.970541983 & SAO 28753 & G9V & 7 & 5308$\pm$36 & 28 & 0.84$\pm$0.03 & 30 & 46.17$\pm$0.02 & (1.62$\pm$0.32)$\times$10$^{-12}$ & 63 & 0.33$\pm$0.10 & 40 \\
300.182122282 & $+$22.709775924 & HD 189733 & K2V & 7 & 5019$\pm$23 & 44 & 0.76$\pm$0.01 & 44 & 50.57$\pm$0.03 & (7.48$\pm$1.50)$\times$10$^{-13}$ & 65 & 6.4$^{+4.8}_{-4.2}$ & 30 \\
085.334752342 & $+$53.478804107 & HR 1925 & K0V & 7 & 5243$\pm$32 & 28 & 0.850$\pm$0.050 & 18 & 81.43$\pm$0.05 & (4.57$\pm$0.91)$\times$10$^{-12}$ & 63 & 0.5$\pm$0.1 & 22 \\ 
063.808266649 & $-$07.667602660 & 40 Eri A & K0.5V & 4 & 5147$\pm$14 & 18 & 0.805$\pm$0.004 & 18 & 198.57$\pm$0.51 & (6.93$\pm$1.39)$\times$10$^{-12}$ & 63 & 11.76$^{+1.92}_{-5.19}$ & 22 \\ 
143.181619543 & $+$26.987467661 & DX Leo & K1V & 4 & 5315$\pm$35 & 28 & 0.81$\pm$0.02 & 15 & 55.32$\pm$0.06 & (2.31$\pm$0.46)$\times$10$^{-12}$ & 63 & 0.25$\pm$0.05 & 43 \\ 
168.637658369 & $+$25.710596159 & HD 97658 & K1 & 7 & 5157$\pm$29 & 28 & 0.72$\pm$0.02 & 30 & 46.35$\pm$0.05 & (9.1$^{+9.9}_{-4.5}$)$\times$10$^{-13}$ & 59 & 9.7$\pm$2.8 & 44 \\ 
271.365865680 & $+$02.494347134 & 70 Oph A & K1 & 1 & 5407$\pm$52 & 18 & 0.831$\pm$0.004 & 18 & 195.22$\pm$0.10 & (2.14$\pm$0.43)$\times$10$^{-11}$ & 63 & 1.3$\pm$0.3 & 37 \\
289.752832267 & $+$41.631884026 & Kepler-444 & K1 & 1 & 5053$\pm$45 & 28 & 0.748$\pm$0.006 & 24 & 27.41$\pm$0.03 & (5.47$\pm$1.77)$\times$10$^{-14}$ & 60 & 11.23$^{+0.91}_{-0.99}$ & 33 \\
043.135621621 & $-$12.770528318 & EP Eri & K1.5V & 4 & 5167$\pm$44 & 21 & 0.79$^{+0.04}_{-0.02}$ & 15 & 96.54$\pm$0.04 & (6.05$\pm$1.21)$\times$10$^{-12}$ & 63 & 0.2$\pm$0.1 & 22 \\ 
\dots & \dots & $\alpha$ Cen B & K2IV & 4 & 5232$\pm$8 & 17 & 0.8761$\pm$0.0029 & 17 & 796.92$\pm$25.90 & (1.50$\pm$0.30)$\times$10$^{-10}$ & 63 & 5.3$\pm$0.3 & 34 \\
004.587759006 & $+$30.955409811 & PW And & K2V & 11 & 4796$\pm$80 & 23 & 0.72$^{+0.01}_{-0.02}$ & 15 & 35.29$\pm$0.05 & (2.31$\pm$0.46)$\times$10$^{-12}$ & 63 & 0.15$^{+0.05}_{-0.02}$ & 45 \\ 
053.228430601 & $-$09.458171515 & $\epsilon$ Eri & K2V & 4 & 5077$\pm$35 & 18 & 0.735$\pm$0.005 & 18 & 312.22$\pm$0.47 & (6.1$\pm$0.2)$\times$10$^{-11}$ & 59 & 0.5$\pm$0.1 & 22 \\ 
057.604601528 & $+$17.246412457 & V471 Tau & K2V & 10 & 5291$\pm$167 & 27 & 0.83$\pm$0.03 & 31 & 20.96$\pm$0.04 & (2.87$\pm$0.57)$\times$10$^{-12}$ & 63 & 0.63$\pm$0.05 & 41 \\ 
143.105446328 & $-$11.184489206 & LQ Hya & K2V & 12 & 5376$\pm$43 & 17 & 1.0029$\pm$0.0158 & 17 & 54.68$\pm$0.07 & (4.01$\pm$0.80)$\times$10$^{-12}$ & 63 & 0.07$^{+0.03}_{-0.02}$ & 43 \\ 
258.834197764 & $-$26.606646978 & 36 Oph A & K2 & 1 & 5103$\pm$29 & 28 & 0.76$\pm$0.02 & 15 & 167.82$\pm$0.16 & (1.42$\pm$0.28)$\times$10$^{-11}$ & 63 & 1.7$\pm$0.4 & 22 \\
349.865586265 & $+$79.003831692 & V368 Cep & K2V & 7 & 5075$\pm$36 & 28 & 0.76$^{+0.02}_{-0.01}$ & 15 & 52.73$\pm$0.03 & (2.99$\pm$0.60)$\times$10$^{-12}$ & 63 & 0.09$^{+0.06}_{-0.04}$ & 43 \\ 
088.517219072 & $-$60.023728916 & HD 40307 & K2.5V & 4 & 4925$\pm$71 & 22 & 0.71$\pm$0.01 & 15 & 77.27$\pm$0.03 & (2.0$^{+2.2}_{-0.9}$)$\times$10$^{-13}$ & 59 & 6.9$\pm$4.0 & 44 \\ 
311.937623890 & $-$36.595010211 & Speedy Mic & K3V & 13 & 4609$\pm$80 & 23 & 1.42$^{+0.03}_{-0.06}$ & 15 & 14.98$\pm$0.27 & (1.85$\pm$0.37)$\times$10$^{-12}$ & 63 & 0.03$\pm$0.01 & 40 \\ 
330.871402102 & $-$56.796902312 & $\epsilon$ Ind & K4V & 4 & 4649$\pm$36 & 21 & 0.732$\pm$0.006 & 32 & 274.81$\pm$0.25 & (3.10$\pm$0.62)$\times$10$^{-11}$ & 63 & 1.6$\pm$0.2 & 21 \\ 
316.747737594 & $+$38.763411107 & 61 Cyg A & K5 & 5 & 4361$\pm$17 & 18 & 0.665$\pm$0.005 & 35 & 285.95$\pm$0.10 & (1.72$\pm$0.34)$\times$10$^{-11}$ & 63 & 6.0$\pm$1.0 & 35 \\
147.782122563 & $-$43.504816171 & HD 85512 & K6V & 4 & 4455$\pm$80 & 23 & 0.70$^{+0.02}_{-0.04}$ & 15 & 88.62$\pm$0.04 & (1.2$^{+1.8}_{-0.5}$)$\times$10$^{-12}$ & 59 & 5.61$\pm$0.61 & 50 \\
077.958661296 & $-$45.043019828 & Kapteyn's Star & sdM1 & 2 & 3570$\pm$80 & 20 & 0.291$\pm$0.025 & 20 & 254.23$\pm$0.04 & (5.27$\pm$0.26)$\times$10$^{-13}$ & 20 & 11.5$^{+0.5}_{-1.5}$ & 20 \\
311.291136883 & $-$31.342450016 & AU Mic & M1 & 6 & 3588$\pm$87 & 16 & 0.698$\pm$0.021 & 16 & 102.83$\pm$0.05 & (1.07$\pm$0.04)$\times$10$^{-11}$ & 61 & 0.02$\pm$0.01 & 45 \\ 
259.751060918 & $-$34.997765110 & GJ 667C & M1.5V & 48 & 3472$\pm$73 & 6 & 0.37$\pm$0.05 & 6 & 138.02$\pm$0.09 & (5.2$\pm$0.9)$\times$10$^{-13}$ & 59 & $>$2 & 51  \\
323.391261617 & $-$49.012516898 & GJ 832 & M2 & 6 & 3522$\pm$60 & 6 & 0.442$\pm$0.013 & 16 & 201.40$\pm$0.04 & (9.5$\pm$0.6)$\times$10$^{-13}$ & 59 & 8.4 & 46 \\ 
070.735385329 & $+$18.953357960 & GJ 176 & M2 & 6 & 3679$\pm$77 & 18 & 0.453$\pm$0.022 & 18 & 105.56$\pm$0.07 & (3.9$\pm$0.2)$\times$10$^{-13}$ & 59 & 4.0$\pm$0.3 & 18 \\ 
154.898887344 & $+$19.869815836 & AD Leo & M3 & 6 & 3308$\pm$84 & 16 & 0.422$\pm$0.013 & 16 & 201.37$\pm$0.07 & (8.07$\pm$0.20)$\times$10$^{-12}$ & 61 & 0.025--0.3 & 36 \\ 
175.550536327 & $+$26.703066902 & GJ 436 & M3 & 6 & 3416$\pm$53 & 18 & 0.455$\pm$0.018 & 18 & 102.50$\pm$0.09 & (2.1$\pm$0.3)$\times$10$^{-13}$ & 59 & 4.2$\pm$0.3 & 18 \\ 
229.856472500 & $-$07.722693495 & GJ 581 & M3 & 6 & 3415$\pm$87 & 16 & 0.330$\pm$0.01 & 16 & 158.75$\pm$0.05 & (1.1$^{+0.3}_{-0.2}$)$\times$10$^{-12}$ & 59 & 4.1$\pm$0.3 & 18 \\
258.831399085 & $+$04.960679473 & GJ 1214 & M4V & 49 & 3008$\pm$96 & 16 & 0.203$\pm$0.008 & 16 & 68.27$\pm$0.17 & (1.3$^{+1.4}_{-0.5}$)$\times$10$^{-14}$ & 59 & 5--10 & 52 \\
343.323973712 & $-$14.266595816 & GJ 876 & M4 & 6 & 3129$\pm$19 & 18 & 0.376$\pm$0.006 & 18 & 213.87$\pm$0.08 & (3.9$\pm$0.4)$\times$10$^{-13}$ & 59 & 9.51$\pm$0.58 & 47 \\ 
153.709069589 & $-$47.154935771 & GJ 1132 & M4 & 1 & 3216$\pm$100 & 16 & 0.217$\pm$0.008 & 16 & 79.25$\pm$0.04 & (2.9$^{+0.4}_{-0.3}$)$\times$10$^{-14}$ & 64 & $>$5.0 & 62 \\
341.702962603 & $+$44.332017083 & EV Lac & M5 & 6 & 3273$\pm$86 & 16 & 0.341$\pm$0.010 & 16 & 198.01$\pm$0.04 & (2.75$\pm$0.55)$\times$10$^{-12}$ & 63 & 0.025--0.3 & 36 \\
217.393465743 & $-$62.676182103 & Prox Cen & M5.5 & 14 & 2840$\pm$88 & 16 & 0.154$\pm$0.005 & 16 & 768.50$\pm$0.20 & (4.37$\pm$0.07)$\times$10$^{-12}$ & 61 & 5.3$\pm$0.3 & 34 \\ 
346.626391870 & $-$05.043461802 & Trappist-1 & M8 & 53 & 2550$\pm$55 & 53 & 0.117$\pm$0.004 & 53 & 80.45$\pm$0.12 & (7.5$\pm$2.3)$\times$10$^{-15}$ & 53 & 7.6 $\pm$2.2 & 54 \\
278.907460881 & $+$32.994892032 & LSR J1835$+$3259 & M8.5 & 55 & 2578$\pm$39 & 56 & 0.137$\pm$0.007 & 57 & 175.82$\pm$0.09 & (7.0$\pm$3.5)$\times$10$^{-15}$ & 58 & \dots & \dots   \\
\enddata
\tablenotetext{a}{All parallaxes come from {\it Gaia} \citep{gaia16, gaia18} with the following exceptions: Procyon, $\alpha$ Cen A, $\alpha$ Cen B, and HR 4657 \citep{van07}, and  Ross 1044 \citep{finch16}.}
\tablenotetext{b}{The \lya\ flux range given for the Sun encompasses both the active and quiet Sun values given in \cite{linsky13}.}
\tablerefs{(1) \cite{bid85}; (2) \cite{haw96}; (3) This work; (4) \cite{gray06}; (5) \cite{alo15}; (6) \cite{gai14}; (7) \cite{gray03}; (8) \cite{cay01}; (9) \cite{gray01}; (10) \cite{hus06}; (11) \cite{zuck04}; (12) \cite{rice98}; (13) \cite{torres06};  (14) \cite{bess91}; (15) \cite{gaia18}; (16) \cite{muir18}; (17) \cite{boy13}; (18) \cite{yee17}; (19) \cite{stass18}; (20) \cite{gui16}; (21) \cite{ram13}; (22) \cite{agu18}; (23) \cite{cas11}; (24) \cite{silva15}; (25) \cite{newt15}; (26) \cite{cum17}; (27) \cite{huber16}; (28) \cite{luck17}; (29) \cite{val05}; (30) \cite{brew16}; (31) \cite{boch18}; (32) \cite{chen17}; (33) \cite{camp15}; (34) \cite{joyce18}; (35) \cite{kerv08}; (36) \cite{shk09}; (37) \cite{ram12}; (38) \cite{pace13}; (39) \cite{eiroa13}; (40) \cite{plav09}; (41) \cite{perry98}; (42) \cite{del16}; (43) \cite{mesh17}; (44) \cite{bon16}; (45) \cite{bell15}; (46) \cite{bry09}; (47) \cite{mann15}; (48) \cite{lurie14}; (49) \cite{newt14}; (50) \cite{pepe11}; (51) \cite{ang13}; (52) \cite{lal14}; (53) \cite{bour17}; (54) \cite{burg17}; (55) \cite{reid03}; (56) \cite{rojas12}; (57) \cite{ditt14}; (58) \cite{saur18}; (59) \cite{young16}; (60) \cite{bour17b}; (61) \cite{young17}; (62) \cite{bert15}; (63) \cite{wood05}; (64) \cite{waalkes19}; (65) \cite{linsky13}; (66) \cite{bah95}  }
\end{deluxetable*}
\end{longrotatetable}

\acknowledgments
This work is based on observations made with the NASA/ESA {\it Hubble Space Telescope}, obtained at the Space Telescope Science Institute, which is operated by the Association of Universities for Research in Astronomy, Inc., under NASA contract NAS 5-26555. These observations are associated with program \#15283. Support for program \#15283 was provided by NASA through a grant from the Space Telescope Science Institute, which is operated by the Association of Universities for Research in Astronomy, Inc., under NASA contract NAS 5-26555.  A.~S. and E.~S. appreciate support from NASA/Habitable Worlds grant NNX16AB62G.  T.~B. was supported in part by NASA HST grant HST-AR-13911 and NASA Habitable Worlds grant NNX16AB62G.This work is based on observations made with the NASA Galaxy Evolution Explorer. {\it GALEX} is operated for NASA by the California Institute of Technology under NASA contract NAS5-98034.  This work has made use of data from the European Space Agency (ESA) mission {\it Gaia} (\url{https://www.cosmos.esa.int/gaia}), processed by the {\it Gaia} Data Processing and Analysis Consortium (DPAC, \url{https://www.cosmos.esa.int/web/gaia/dpac/consortium}). Funding for the DPAC has been provided by national institutions, in particular the institutions participating in the {\it Gaia} Multilateral Agreement.

\software{{\tt\string emcee} \citep{fm13}}

\end{document}